\documentstyle[aps]{revtex}
\begin{document}
\author{LiXiang Cen$^1$, XinQi Li$^1$ and YiJing Yan$^2$}
\address{$^{1}$NLSM, Institute of Semiconductors, The Chinese Academy of Sciences,\\
Beijing 100083, P.R. China, \\
$^{2}$Department of Chemistry, Hong Kong University of Science and\\
Technology, Kowloon, Hong Kong}
\title{Comment on ``Nonadiabatic Conditional Geometric Phase Shift with NMR''}
\maketitle


Realizing quantum computation by means of geometric origin (adiabatic cyclic
Berry phase) is now receiving considerable attention due to its intrinsic
tolerance to noise. In a recent letter, Wang and Keiji \cite{wang} explored
the nonadiabatic implementation of the geometrical quantum phase shift in a
nuclear magnetic resonance (NMR) system that previously proposed in Ref. 
\cite{nature,kert}. However, it should be further clarified that a parallel
extension of the adiabatic scenario, i.e., the multi-loop operation sequence
of Eq. (11) in Ref. \cite{wang}, cannot realize such a goal, even if the
resonant case is concerned \cite{eprint}.

Differing from the adiabatic scheme, to keep up the eigenstate of $H_0$ with
the speedy rotating fields, two opposite vertical fields are needed as
performing the twice opposite cyclic evolutions, i.e., the 
({\tiny $\! 
\begin{array}{l}
\omega _z \\ 
C
\end{array}
\!\!$}) and ({\tiny $\! 
\begin{array}{l}
{\bar{\omega}}_z \\ 
{\bar{C}}
\end{array}
\!\!$}) 
in the notation of Ref. \cite{wang}, respectively. It is not difficult to
show that, for a fixed recurrent initial state, these two cyclic evolutions
induce the same total phase (the Lewis-Riesenfeld phase \cite{lewi}): $\phi
_C=\phi _{\bar{C}}$. Consequently, one can verify that the four-loop
operation sequence, Eq.\ (11) of Ref.\ \onlinecite{wang}, provides nothing
but an exactly identical total phase for the four computational bases \{$%
|\uparrow \uparrow \rangle ,|\uparrow \downarrow \rangle ,|\downarrow
\uparrow \rangle ,|\downarrow \downarrow \rangle $\}. It is to say that such
an operation sequence, though appearing to be a natural extension of the
adiabatic version, in fact is of no use for geometrical quantum computation.

The origin causing the failure of such an attempt can be revealed in a more
profound manner. The key idea to obtain conditional geometric phases in the
previous propositions in an NMR system [2,3] uses the fact that the cyclic
adiabatic evolutions in opposite directions induce the same dynamical phase
and negative geometric phases; so that in the whole procedure the dynamical
phases accumulated for different bases are identical which can thus be
eliminated as a global phase and the only retained geometric phases are
different for the four bases which implies the conditional geometric shift.
Now consider the nonadiabatic extension of such a version. The characterized
time-dependent Hamiltonian system, i.e.,\ the spin-half nucleus in a
rotating magnetic field, has an invariant, $I(t)$: 
\begin{equation}
\frac{dI(t)}{dt}=\frac{\partial I(t)}{\partial t}-i[I(t),H(t)]=0,
\label{equa}
\end{equation}
and the eigenvectors of $I(t)$ form the recurrent solutions of the system.
In detail, $I(t)$ can be calculated conveniently from the algebraic
dynamical method\cite{gauge}. For the case of a constantly rotating magnetic
field, it is given by 
\begin{equation}
I(t)=H(t)-\frac{\omega }{2}{\mbox{\boldmath $n$}}\cdot {%
\mbox{\boldmath
$\sigma$}},  \label{invar}
\end{equation}
where \mbox{\boldmath $n$} stands for the rotation direction and $\omega $
the rotation magnitude. According to algebraic dynamics \cite{gauge,geome},
the invariant $I(t)$ is related to the total phase of the wave function and
the fraction of it [i.e., the second term of Eq.\ (\ref{invar})] indicates a
gauge potential which is related to the nonadiabatic Berry phase. Note that
the two opposite cyclic Hamiltonians that satisfy $H_{C}(t)=H_{\bar{C}}(T-t)$
induce different instantaneous gauge potentials and thus lead to different
invariants: $I_{C}(t)\neq I_{\bar{C}}(T-t)$. In comparison with the
adiabatic situation, the deviation caused here is twofold: the recurrent
solutions of the two opposite nonadiabatic evolutions become different, and
as a result, the dynamical phases, defined in terms of the expected value of
Hamiltonian over the eigenstate of the invariant $I(t)$, induced by the two
opposite processes are also different in general. In a sense, the latter
deviation is fatal for geometrical realization of quantum computation. The
four-loop operation sequence proposed by Wang and Keiji does arrive at the
same recurrent instantaneous states for the evolutions 
({\tiny $\! 
\begin{array}{l}
\omega _{z} \\ 
C
\end{array}
\!\!$}) and ({\tiny $\! 
\begin{array}{l}
{\bar{\omega}}_{z} \\ 
{\bar{C}}
\end{array}
\!\!$}), 
since the scheme uses two different Hamiltonians that correspond to an
identical invariant (hence they induce the same total phase as was pointed
out in the previous paragraph). However, the dynamical phases induced by the
two evolutions are different and thus they cannot be removed via the
proposed scheme \cite{note}.

Note added: This work was finished in January 2002 and submitted to PRL.
Very recently, we noticed that the same misunderstanding occurred in the
paper by Zhu etc. [Phys. Rev. Lett. 89, 097902 (2002)]. That is, the authors
neglected the fact that the recurrent solutions of the opposite nonadiabatic
evolutions are distinctly different, and the dynamical phases induced
accordingly are also different and cannot be removed via the current
schemes. The assertion that their scheme can implement nonadiabatic
geometrical quantum computation thus is invalid.

\acknowledgements

This work was supported in part by the Postdoctoral Science Foundation, the
special funds for Major State Basic Research Project No. G001CB3095 of
China, and the Research Grants Council of the Hong Kong Government.


\begin{references}
\bibitem{eprint}  X.-B. Wang and M. Keiji, e-print quant-ph/0101038. In this
amended manuscript, the authors believed that the resonant case of the the
multi-loop scheme can realize the geometric quantum phase shift. However, we
will show that it is not the case.
\bibitem{note}  For a resonant case, the recurrent states are in the
vertical of the rotating axis and the dynamical phases induced by the two
opposite cyclic evolutions happen to be identical. However, the geometric
phase induced now becomes exactly zero in view that the gauge potential is
antiparallel to the rotating axis. One can also verify that this is a
natural consequence in accord with the result described in the second
paragraph.
\end{references}
\end{document}